# Electro-optically controlled divided-pulse amplification

**HENNING STARK,**[1,*] **MICHAEL MÜLLER,**[1] **MARCO KIENEL,**[1,2,4] **ARNO KLENKE,**[1,2] **JENS LIMPERT,**[1,2,3] **AND ANDREAS TÜNNERMANN**[1,2,3]

[1] *Institute of Applied Physics, Abbe Center of Photonics, Friedrich-Schiller-Universität Jena, Albert-Einstein-Straße 15, 07745 Jena, Germany*
[2] *Helmholtz-Institute Jena, Fröbelstieg 3, 07743 Jena, Germany*
[3] *Fraunhofer Institute for Applied Optics and Precision Engineering, Albert-Einstein-Str. 7, 07745 Jena, Germany*
[4] *Currently with Active Fiber Systems GmbH, Wildenbruchstraße 15, 07745 Jena, Germany*
\* *lars.henning.stark@uni-jena.de*

**Abstract:** A novel technique for divided-pulse amplification is presented in a proof-of-principle experiment. A pulse burst, cut out of the pulse train of a mode-locked oscillator, is amplified and temporally combined into a single pulse. High combination efficiency and excellent pulse contrast are demonstrated. The system is mostly fiber-coupled, enabling a high interferometric stability. This approach provides access to the amplitude and phase of the individual pulses in the burst to be amplified, potentially allowing the compensation of gain saturation and nonlinear phase mismatches within the burst. Therefore, this technique enables the scaling of the peak power and pulse energy of pulsed laser systems beyond currently prevailing limitations.

**OCIS codes** (140.3298) Laser beam combining; (320.7090) Ultrafast lasers; (140.3510) Lasers, fiber; (140.3280) Laser amplifiers.

## 1. Introduction

Ultrafast high-power lasers pose unique and indispensable tools for a plethora of medical, industrial and scientific applications and open up, due to their consistently increasing brightness, ever new application fields such as, for instance, wakefield particle acceleration [1] and high-harmonic generation [2]. However, further brightness scaling is hampered by physical and technological challenges. For example, high pulse peak intensities induce detrimental nonlinear effects such as, for instance, self-phase modulation and self-focusing, impairing the pulse and beam quality and, ultimately, resulting in optical damage of the media involved. One of today's standard techniques to mitigate nonlinear effects, chirped-pulse amplification (CPA) [3], allows reaching formerly unattainable peak powers, but any further scaling is limited by the grating size available. In the case of fiber lasers, the mode-area scaling of small-core high-NA fibers to very-large-mode-area low-NA fibers [4] allowed for a tremendous increase of peak power. However, as with CPA, a further power scaling following this approach is limited, mostly due to the tight manufacturing tolerances required to make such fibers. Therefore, novel approaches for further peak power scaling are needed; a prominent example of them being coherent beam combining (CBC) [5,6]. This technique exploits parallelization of the amplification process with multiple amplifiers. The subsequent coherent superposition of all the beams from these amplifiers leads to a significant pulse peak power and average power enhancement. For this approach, especially fiber lasers have proven their suitability, since their simple architecture allows for compact and stable setups, while their reproducible and outstanding beam quality enables a highly efficient combining process.

The amplification of temporally separated pulses and their subsequent combination into a single pulse can be interpreted as the time-domain counterpart of CBC. Thus, this technique allows mitigating nonlinear effects in the optical medium. Regarding this strategy, besides cavity enhancement [7,8] and coherent pulse stacking [9], divided-pulse amplification (DPA) [10-13] has especially shown its potential. In DPA, the temporally separated pulses are usually generated from and stacked in a series of optical delay lines. Using actively controlled DPA (ADPA) [14], where the setups for pulse division and combining are separated from one another, in conjunction with CBC, an amplification of four-pulse bursts with subsequent combining generated pulses with an energy of 12 mJ and a duration of 262 fs, which corresponds to 35 GW peak power [15]. Still, in this experiment, the energy-scaling potential of the individual fibers was not fully harvested, since a significant part of the energy stored in the fibers was not extracted. However, a further extraction of this residual energy by means of an increase of the number of pulses in the burst is hampered. In ADPA, the amount and sizes of the optical delay lines make the setup significantly more complex and sophisticated. Additionally, gain saturation occurs (since a considerable part of the energy stored in the active fiber is extracted), leading to the individual pulses of the burst experiencing different amplification levels. This, in turn, results in a mismatch of the amplitude and nonlinear phases among the pulses, which impairs their subsequent combination. In ADPA amplitude pre-shaping allows compensating for amplitude and also phase mismatches to a certain extent [16]. However, since in this approach amplitude and nonlinear phase cannot be shaped independently so far, a residual phase and/or amplitude mismatch remains, substantially reducing the overall performance of the combining. Therefore, the key to scale DPA to even more pulse replicas lies in an extensive control of the amplitude and phase of the individual pulses and in a reduction of the size and complexity of the optical setup.

In this contribution, a novel concept for creating and amplifying a burst of pulses (as a power of two) is introduced. This technique allows stacking the pulses in an arbitrarily long series of optical delay lines. The approach potentially allows for an independent and individual control of amplitude and phase of each pulse while significantly reducing the complexity of the optical setup as it is largely fiber-coupled. We call this approach electro-optically controlled divided-pulse amplification (EDPA) as its key components are electro-optic phase modulators (EOM).

This paper is organized as follows. First, the fundamental principle of EDPA is described. Then, the setup of this proof-of-principle experiment is explained. Next, the experimental results are presented and discussed. Finally, a summary of the experimental findings is given, together with an outlook of the application of EDPA to high-power ultrafast lasers.

## 2. Electro-optically controlled divided-pulse amplification

In classic DPA approaches, a single pulse is temporally split up, by a set of $N$ delay lines, into a burst of $2^N$ pulses, which leads to a lower pulse peak intensity in the laser amplifiers. For the subsequent temporal recombination with delay lines, and, therefore, an increased peak power in the end, the pulse burst has to have a distinctive polarization and phase pattern in order to send the individual pulse replicas into their corresponding delay path. The necessary pattern can be found by propagating a single pulse backwards through the combining stage, as depicted exemplary for four pulses in Fig. 1 (from right to left). Here, the single pulse on the right is initially linearly polarized in a 45° angle, i.e. it comprises equal amounts of p- and s-polarization. Propagating the pulse to the left, the polarizing beam splitter (PBS) 2 transmits the p-polarized component while the s-polarized component is reflected into the delay line (DL) 2. Therefore, the s-polarized part is delayed and, finally, reflected a second time by PBS 2, lining up the s-polarized component behind the previously transmitted p-polarized component. Since both pulse replicas are orthogonally polarized, a rotation by 45° with a half-wave plate (HWP) allows separating their p- and s-polarized components again by means of PBS 1, sending the s-polarized parts into DL 1. In this case, DL 1 has half the length of the previously passed DL 2. Consequently, a burst of four pulses with alternating orthogonal polarizations and a distinctive phase pattern is generated. In reverse, a pulse burst with this pattern going through the combining stage (from left to right in Fig. 1) will be combined into a single pulse. There are several approaches to generate this pattern. For instance, in passive DPA [13], the polarization pattern is generated by the very same delay line arrangement which is subsequently used for combining. In ADPA [14], on the other hand, the division and combining stages are separated, requiring a second set of delay lines.

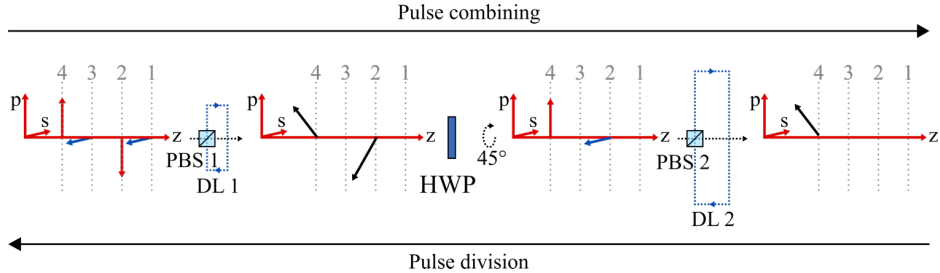

Fig. 1. Simplified representation of the combination of four pulses (when going from left to right). The same setup can be used for pulse division when going from right to left. In this latter case, the rotation by the half-wave plate (HWP) and the propagation in the delay lines (DL) are reversed. (PBS: polarizing beam splitter.)

EDPA employs an identical setup for the pulse combination as conventional DPA. However, the pulse burst with the specific polarization pattern is generated without any delay lines in a compact fiber-integrated front-end. The principle of the pattern generation is illustrated in Fig. 2 (from left to right) once again for four pulses. It starts with a p-polarized burst of equal pulses, which can be cut out directly from the pulse train emitted by an oscillator by means of an amplitude modulator. The burst is then spatially split in two equal parts that are sent to two amplifier channels, each containing a phase controlling element. Here, distinctive patterns of the relative phases are imprinted on both bursts. These phase patterns eventually will produce the polarization pattern required for the temporal combination and consist, in this idealized case, of relative phases of 0 and $\pi$. Since both channels contain amplifiers, this allows for a temporally and spatially separated amplification. Thereafter, the polarization of one channel is rotated by 90°, so that one channel is p-polarized and the other one is s-polarized. The interferometric superposition of both channels leads to a burst of pulses with, in this case, an alternating orthogonal polarization pattern. Next, the polarization is rotated by 45° degrees, so that the individual pulses of the generated burst are either p- or s-polarized. Due to the chosen phase patterns, this polarization pattern is exactly the one required for the combining stage that has been described above. Analogous to the case of the polarization pattern, the phase patterns that have to be imprinted on the pulses are found by propagating the polarization pattern in reverse (i.e. from right to left) through the setup shown in Fig. 2. It is important to note that this intuitive technique of generating the necessary polarization pattern and, consequently, the required phase patterns by propagating a pulse backwards through the system can be applied to any arbitrary number of optical delay lines.

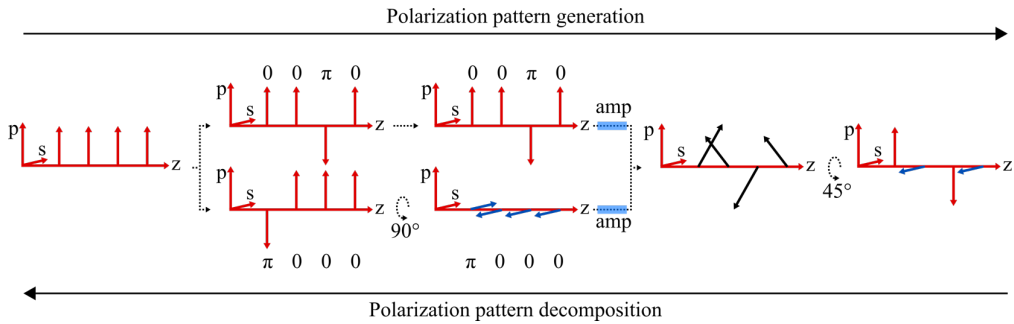

Fig. 2. Generation of the polarization and phase pattern (from left to right). The individual steps are: splitting of the burst and imprinting the phase patterns; rotating one channel by 90°; amplifying and superposing both channels; rotating the polarization by 45°. In case of pattern decomposition (from right to left) the polarization rotations are reversed.

In conclusion, EDPA uses an interferometric superposition of two orthogonally polarized pulse bursts with distinctive phase patterns to generate the polarization pattern which is required for the temporal combination with a set of delay lines. The corresponding phase patterns, which need to be imprinted, can be found by propagating a single pulse backwards through the whole system to the phase modulators.

As soon as gain saturation occurs in the amplifiers, some amplitude and nonlinear phase mismatches between the individual pulses appear and they reduce the performance of the combining process. This is where EDPA shows its full potential, since it allows for compensating the amplitude mismatch for any arbitrary number of pulses by pre-shaping the burst with the amplitude modulator. Furthermore, EDPA has the potential to significantly reduce the impact of the nonlinear phase mismatch, since it provides access to the phase of every individual pulse in the burst. To do this, in theory, the phase modulators simply have to add extra phases, individually customized to the demands of every single pulse, to the phases 0 and $\pi$ of the phase patterns used in the idealized case without saturation. This pre-compensates variations of the accumulated nonlinear phase between the pulses in the subsequent amplification and substantially increases the performance of the combination. As another key feature, the number of pulses to be combined can be easily and arbitrarily scaled (but the final number of pulses always has to be a power of two) by changing the duration of the transmission window of the amplitude modulator and adjusting the phase patterns imprinted by the phase modulators (provided that the required delay lines are included in the combining stage). Furthermore, the extensive amplitude and phase corrections can be freely added to any desired number of pulses, posing an essential enhancement of former DPA techniques in the generation of high-power ultrafast laser systems.

In this proof-of-principle experiment, however, only the general functional capability and performance of this technique are investigated, leaving the examination of phase and amplitude pre-compensation to future high-power experiments.

## 3. Experimental setup

The setup of the EDPA experiments used for the combination of four temporally separated pulses is schematically depicted in Fig. 3. It mainly consists of two parts: a fiber-coupled front-end (using PM980 fibers) and a free-space combining stage. At the beginning of the master oscillator power amplifier (MOPA) architecture, a homebuilt mode-locked fiber oscillator with a repetition rate of $f_\text{rep}$=108 MHz emits pulses with a duration of 190 ps at a center wavelength of 1030 nm. A synchronous countdown counts the generated laser pulses with the help of a photo diode (PD). This is used to set the temporal distance between two subsequent pulse bursts, which is the inverse of the burst repetition rate $f_\text{burst}$. For this, the synchronous countdown emits an electric signal every $N_\text{SC}$ laser pulses, which is referred to as the clock signal, since it determines the timing of all electronic devices to be used. An arbitrary waveform generator (AWG) is used to drive an acousto-optic modulator (AOM) that applies a 37 ns transmission window to the pulse train coming from the oscillator. Thus, bursts of four pulses are created with the burst repetition rate $f_\text{burst} = f_\text{rep} / N_\text{SC}$. Next, the resulting laser signal is spatially divided into two channels. Each of them contains a fiber-coupled EOM, both driven by a two-channel AWG with a sampling rate of 1.2 GS/s.

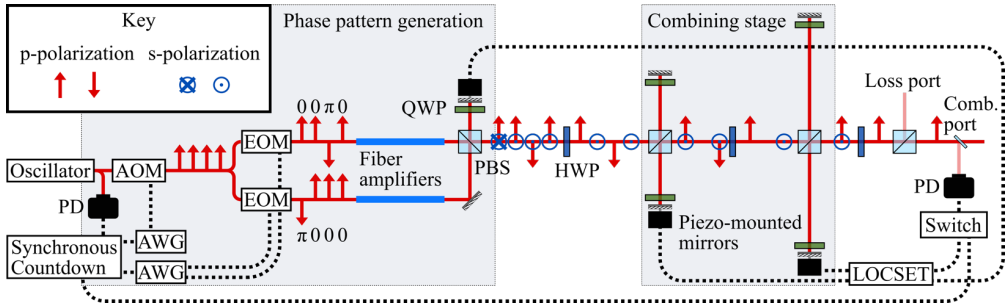

Fig. 3. Schematic illustration of the EDPA setup used for the combination of four temporally separated pulses from two channels. (PD: photo diode, AOM: acousto-optic modulator, AWG: arbitrary waveform generator, EOM: electro-optic modulator, QWP: quarter-wave plate, HWP: half-wave plate, PBS: polarizing beam splitter.)

The EOMs add the phase patterns to the pulse bursts in each channel. In the simple case of a negligible nonlinear phase mismatch between the pulses, the phase pattern is set to consist of 0 and π. However, as soon as the nonlinear phase mismatch shows a noticeable impact on the combining, phase adjustments can be freely added to the phase patterns by means of the AWG. One ytterbium-doped polarization-maintaining 6 µm-fiber amplifier per channel amplifies the signals to up to 300 mW each. The signals are coupled out of the fibers, being collimated by aspheric lenses with a focal length of 8 mm. Next, both beams, which are initially p-polarized, pass a PBS with orthogonal propagation directions. On the one side, a 0° mirror with a quarter-wave plate (QWP) in front of it rotates the signal from the one channel to s-polarization and reflects it back to the PBS. The lengths of both channels are interferometrically matched, such that the spatial combination of the two beams results in a pulse burst pattern with alternating orthogonal polarizations. A telescope is applied to increase the beam diameter to approximately 5 mm, reducing the impact of divergence in the long delay lines used for pulse stacking by increasing the Rayleigh length to about 19 m. With a HWP the polarization of the generated pulse burst is rotated by 45°, finally resulting in the same polarization and phase pattern as in conventional DPA. It consists of alternating s- and p-polarized pulses which allows for a subsequent temporal combination.

The combining stage begins with a PBS, which reflects all s-polarized pulses into a delay line, while the p-polarized ones simply go through it. This first delay line has a length of $l_1 = c / f_{rep} = 2.76$ m, determined by the fundamental repetition rate $f_{rep}$ of the master oscillator and the speed of light in air $c$. With this delay line, the delayed pulses (the first and the third one) are coherently combined with their particular subsequent neighbors (the second and the fourth pulse, respectively). A 4f-arrangement is set up in the delay line to further reduce beam divergence and, therefore, maximize the spatial overlap in the combination. Due to the phase patterns imprinted with the EOMs, the polarization pattern after this first temporal combining step again consists of orthogonally polarized pulses. Thus, after a polarization rotation by 45° with a HWP, the pulses are again alternatingly s-polarized and p-polarized. The second delay line has twice the length of the first one, that is $l_2 = 2 \cdot l_1 = 5.52$ m, allowing for a temporal combination into a single pulse. After another rotation with a HWP, the combined pulse is transmitted by another PBS, which, for the most part, separates the successfully combined from the uncombined radiation.

Since the overall setup is an interferometer with multiple arms, an active phase stabilization is required. In this setup, the technique of locking of optical coherence by single-detector electronic-frequency tagging (LOCSET) [17] was used in conjunction with piezo-mounted mirrors in every combining step, modulating distinctive frequencies onto the combined signal. A fraction of the combined signal is detected by a photodiode (PD) as required for the closed-loop control. A demodulation at the respective frequencies delivers the required information to calculate the phase corrections that have to be applied with the piezo-mounted mirrors. However, since in spatio-temporal combining multiple stable states can be phase locked [18], a fast electronic switch is implemented. This introduces a short temporal window in the acquisition of the error signal around the temporal position of the correctly combined pulse. This way, the phase locking of unintended secondary stable states is avoided.

## 4. Experimental results

The quality of the spatial and temporal combination processes is crucial in order to obtain the maximum achievable performance of the corresponding system. Therefore, as a figure of merit, the combining efficiency

$$\eta_{comb} = \frac{E_{comb}}{E_{loss} + E_{comb}} \quad (1)$$

is introduced. It depends on the energy of the successfully combined pulse $E_{comb}$ and the uncombined energy $E_{loss}$.

Another characteristic value, which additionally includes information about the loss inside the combining system due to losses at the optical elements $\eta_{OE} < 1$, is the system efficiency

$$\eta_{sys} = \eta_{comb} \cdot \eta_{OE} = \frac{E_{comb}}{E_{tot}}. \quad (2)$$

This parameter is defined as the ratio between the energy of the combined pulse $E_{comb}$ and the total energy of the amplified pulse burst before the temporal combining stage, represented by

$E_{tot}$. The temporal efficiency $\eta_{temp}$ of the combining is the energy contained in the temporal slot that corresponds to the successfully combined pulse divided by the total energy in the combined signal, which, therefore, also includes pre- and post-pulses. As confirmed later in detail by investigating the temporal contrast, which is in direct relation to the temporal efficiency, the temporal efficiency in EDPA is found to be $\eta_{temp} \approx 1$. Therefore, the measurements of the system efficiency and combining efficiency allow using the corresponding average powers $P_{comb}$, $P_{loss}$ and $P_{tot}$ instead of the energies.

During the experiments, the pulse burst repetition rate is varied while the average power of the signal seeding the main amplifiers is held constant by readjusting the pump power of the pre-amplifier. This results in a change of the total energy and, therefore, of the energy of the combined pulse. However, due to the limited operating range of the active stabilization and the driving EOM electronics, only repetition rates from 135 kHz to 1529 kHz can be applied. The achieved total power is in the range between 455 mW and 562 mW. Consequently, when combining four temporally separated pulses, total energies between 0.35 µJ and 3.5 µJ are generated. The efficiency measurements show consistently high values for all repetition rates applied. Although the theoretically estimated saturation energy of 14 µJ per channel was not reached, considerable signs of saturation were observed which can be seen in the progressively decreasing amplitudes of the individual pulses within the burst as illustrated in Fig. 4(a). In spite of these signs of saturation, there were no noticeable impairing effects on the combining apart from the need to readjust the HWPs in the combining stage. The detailed results of the combining efficiency measurements are shown in blue as a function of the total energy in Fig. 4(b). The values range from 92.5 % to 95.4 %. The system efficiency, colored red, is also included in Fig. 4(b). It shows a similar behavior and it settles between 78.1 % and 82.7 %. The discrepancy between both efficiencies originates from losses at the optical elements. This assumption is confirmed in a separate measurement, showing power losses of the laser beam of up to 13% during a simple propagation through the delay lines. Such power losses are reasonable, since many standard HR mirrors are used in the folded delay lines and the contrast of the PBSs and the accuracy of the wave-plates are limited.

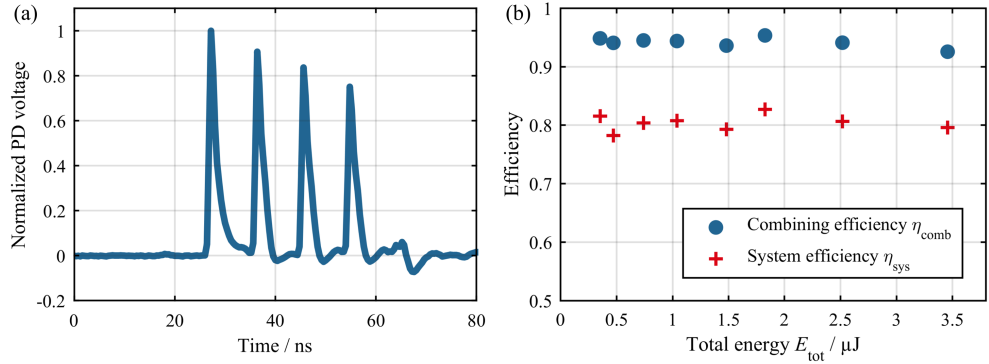

Fig. 4. (a) PD trace of a pulse burst at a total energy of $E_{tot}$ = 3.4 µJ at a burst repetition rate of $f_{burst}$ = 135 kHz before temporal combining. (b) Experimental result of the combining efficiency and system efficiency measurements for the temporal combining of four pulses.

As another figure of merit, the temporal contrast

$$C = 10 \cdot \log_{10} \frac{E_{comb}}{E_{pre}} \qquad (3)$$

is introduced. $E_{comb}$ and $E_{pre}$ denote the energy of the combined pulse and the energy of the strongest pre-pulse, respectively. Both can be seen in the PD voltage trace in Fig. 5(a). For this measurement, only the most powerful pre-pulse is of concern. This is, because, on the one hand, for most applications only the pre-pulse contrast is of interest, since an early impact might disturb the targeted object before the main pulse arrives. On the other hand, the signal is almost symmetric with respect to the combined pulse when the combining efficiency is optimized. Therefore, a measurement of the post-pulse contrast would provide similar values.

The contrast is evaluated from the photo diode (PD) voltage traces of the combined signal, as depicted in Fig. 5(a). As the contrast is high, the peak voltage of the main pulse is measured with neutral density (ND) filters, while the pre-pulse is recorded without any attenuation. Therefore, similar PD voltages can be ensured, eliminating the possibility of falsified measurement data due to PD saturation. Subsequently, by determining the precise attenuation of every single ND filter, the actual contrast is calculated. In Fig. 5(b), the result is depicted as a function of the total energy, showing a temporal contrast in the range of 23.2 dB to 27.3 dB. Since other pre-pulses are negligibly small, the achieved temporal contrast corresponds to a temporal efficiency of $\eta_{temp} > 0.99$, even though a post-pulse that is similar to the pre-pulse is included. This confirms the earlier assumption of $\eta_{temp} \approx 1$.

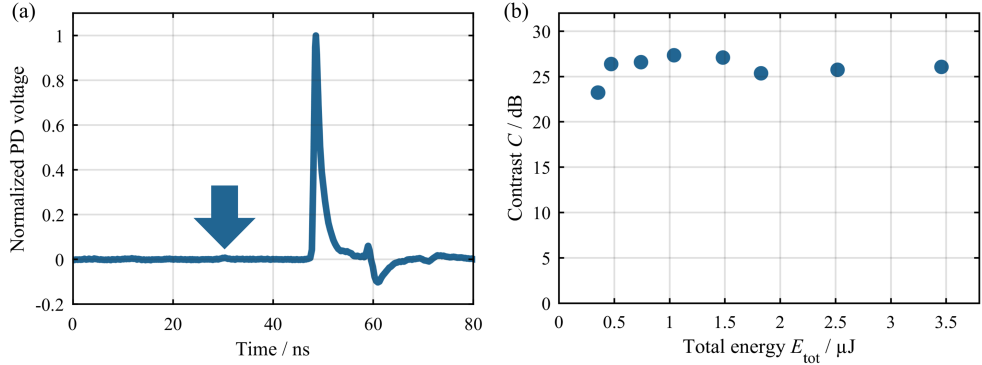

Fig. 5. (a) PD voltage trace of the combined signal from EDPA with four pulses, $E_{tot} = 3.5$ µJ and $f_{burst} = 135$ kHz. The blue arrow highlights the strongest pre-pulse. The visible distortion at about 60 ns has been proven to be a measurement artifact originating from a reflection of the electric signal of the main pulse (ringing). (b) Experimental result of the temporal contrast measurements for the temporal combining of four pulses using EDPA.

For a theoretical estimation of the achievable contrast, Eq. (3) is used. The highest possible $E_{pre}$ originates from the last combining step. Before this step, two pulses remain, but only the first one, having an energy of almost exactly ½·$E_{comb}$, contributes to the pre-pulse. Since the specification of the transmissivity of the PBSs of s-polarized light is given by $T_s < 0.005$, a worst-case value $T_s = 0.005$ is assumed, leaving a pre-pulse with an energy of 0.005·½·$E_{comb}$. This pre-pulse is polarization-rotated by the following HWP and its energy is cut in half by the last PBS, dumping one half of the energy on the loss port. Consequently, the resulting pre-pulse energy is $E_{pre} = $ ½·0.005·½·$E_{comb}$, which can be applied to Eq. (3), resulting in

$$C_{max} = 10 \cdot \log_{10} \frac{2 \cdot 2}{0.005} \approx 29 \text{dB} . \quad (4)$$

The measured contrast is high but still slightly lower than this theoretical maximum, indicating that small combining losses due to, for instance, beam size mismatches or wave front distortions exist. Furthermore, the phase jitter introduced by LOCSET also leads to a small decrease of the temporal contrast on average. However, by optimizing the control parameters of LOCSET, its influence on the temporal contrast can be minimized.

Finally, the stability of the system is investigated. For this, the power of the combined signal resulting from the individual combining steps is temporally resolved with a PD. The signal is then filtered by a 500 kHz low-pass filter, recorded with an oscilloscope and Fourier transformed using the Hann window. The resulting power spectral densities (PSD) are depicted in Fig. 6(a) along with the corresponding integrated power spectral densities in Fig. 6(b). For the temporal combining of four pulses a low relative intensity noise (RIN) of 0.54 % is achieved, while the RIN of a single channel and using exclusively spatial combining is lower at approximately 0.2 %, as expected. The disturbances in the upper kHz regime of the four-pulse combining, which manifest as distinct steps in the integrated PSD, can be assigned to the LOCSET jitter frequencies, which are located at 5 kHz, 6 kHz and 9 kHz. Therefore, in the integrated PSD of the spatial combining, only a single step occurs at 5 kHz, since 6 kHz and 9 kHz were the jitter frequencies for the temporal combining. The edge at 80 Hz is most probably a mechanical resonance caused by the cooling fans of the electric control devices.

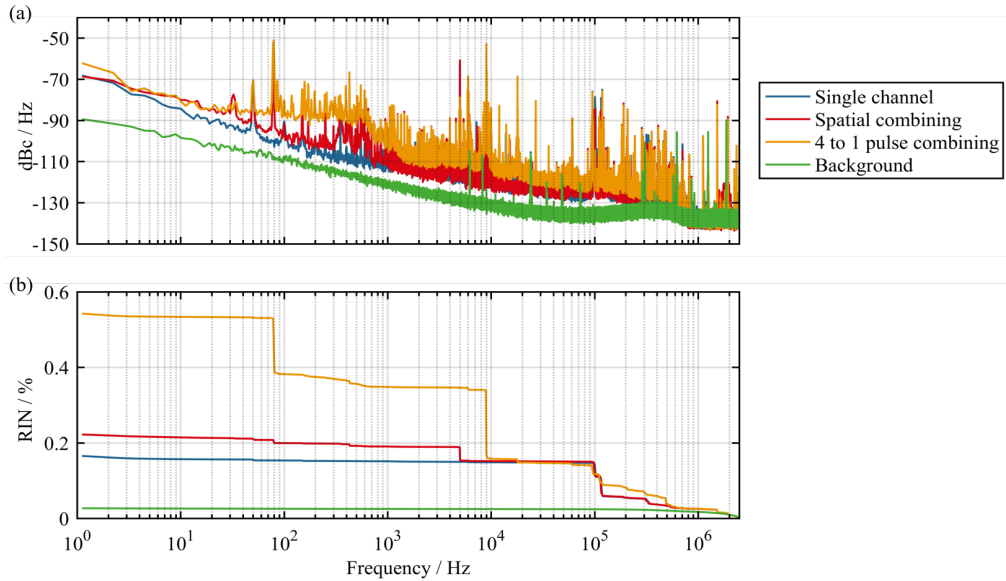

Fig. 6. Power spectral density (a) and integrated power spectral density (b) measured for different combining steps (spatial combining in red and temporal combining of four pulses from two channels each in yellow), a single channel before combining (colored blue) and the background (green) in EDPA at $f_{burst}$=1529 kHz.

The combining of four temporally separated pulses from two channels was analyzed thoroughly and proves that EDPA is performing well. Therefore, the scalability of EDPA is investigated by doubling the amount of pulses to be combined.

For the combining of eight pulses different adjustments and extension need to be made. To generate the pulse burst, the duration of the transmission window of the AOM is doubled to 74 ns. In accordance with Fig. 3, a third delay line is added to the end of the combining stage, between the former last delay line and the last PBS which separates the combined output from the loss port. The new delay line has a length of 11.04 m, which is twice the length of the longest delay line from the four-pulse combining. It is folded by 9 mirrors in 16 reflections and has three subsequent 4f-arrangements set up inside to counteract beam divergence. The phase patterns applied by the two EOMs and, consequently, the generated polarization pattern required for the combining, are determined again by propagating a single pulse backwards through the setup, as explained in Sec. 2 and depicted in Fig. 1 and Fig. 2.

Doubling the number of pulses barely shows any influence on the handling of EDPA. Therefore, with the mentioned adjustments, the scaling of the pulse number in EDPA is relatively straight-forward. Already at the first try, the eight pulses were combined successfully at a fixed pulse burst repetition rate of $f_{burst}$=1075 kHz. Few optimizations of the temporal and spatial alignment in the optical combining setup directly lead to excellent results. A combining efficiency of 89.7 % is reached, which is just barely lower than the result from the four-pulse combining. However, this small decrease of the combining efficiency indicates that there is still potential for improvement, especially in the last delay line. For instance, aberrations and beam parameter mismatches due to imperfect adjustments of the imaging systems and a limited choice of lenses decrease the constructive beam overlap, finally reducing the performance of the combining process. Furthermore, the large number of optical elements in the delay line creates many possibilities for phase front errors.

The measurement of the corresponding system efficiency shows a high value of 76.8 %, which is consistent with the slight reduction of the combining efficiency. As before, the power losses are expected to originate from the large amount of imperfect optical elements. In a last measurement of the eight-pulse combining, the photo diode voltage trace of the combined signal is investigated. Again, the recording shows a temporal contrast with similar results as those seen in the experiments with the combining of four temporally separated pulses.

## 5. Conclusion and prospect

In this contribution, a novel technique for temporal pulse division, subsequent amplification and combination is presented. In a proof-of-principle experiment, four temporally separated pulses from two channels each are combined into a single pulse. The corresponding investigations show high combining efficiencies of up to 95.4 % and good system efficiencies of up to 82.7 %. A high interferometric stability with a measured RIN of 0.54 % together with a high temporal contrast of up to 27.3 dB is achieved, being close to the estimated theoretical optimum. The scalability of this technique is investigated by combining eight temporally separated pulses. A combining efficiency of 89.7 % and a system efficiency of 76.8 % are measured. The complexity of the optical setup as well as the required free-space propagation distance is reduced drastically in comparison with similarly performing ADPA setups. In addition, without any changes to the optical setup, EDPA already provides the possibility for amplitude pre-shaping of pulse bursts with arbitrary lengths to compensate for gain saturation. Furthermore, the EOMs offer access to the phase of every individual pulse. In conventional DPA nonlinear phase mismatches from gain saturation remain despite amplitude pre-shaping. Therefore, phase control poses a powerful tool in all types of DPA, since it potentially allows for an active compensation of the accumulated nonlinear phases and, therefore, a matching of the phases between the individual pulses. In theory, the EOMs also allow adding the phase jitter to the individual pulses, which is required for the active stabilization with LOCSET. This will again simplify the optical setup and probably further improve the performance of EDPA, since the jitter will not need to be applied mechanically with piezo mounted mirrors anymore.

Few issues remain, however, such as, for instance, the considerable losses at the optical elements or the aberrations caused by imperfect imaging in the delay lines. A solution to this will be given by the application of multipass cells, e.g. Herriott type ones [19], with customized high-efficiency concave mirrors. Additionally, highly stable mirror mounts and tailored housings are expected to further improve the temporal stability. These upgrades will also have a positive influence on the temporal contrast, which will be increased even more in future experiments by replacing the PBSs with high-contrast thin film polarizers.

In conclusion, EDPA shows an excellent efficiency, contrast and stability. In conjunction with the significantly reduced size and complexity of the optical setup and the potential for extensive amplitude and phase shaping, EDPA is a promising technique for the generation and further scaling of high peak power radiation with ultrafast laser systems.

**Funding**

Free State of Thuringia (2015FE9158) "PARALLAS", co-funded by the European Union within the framework of the European Regional Development Fund (ERDF); European Research Council (ERC) (617173, 670557).